\documentclass[]{interact}

\usepackage[caption=false]{subfig}

\usepackage[numbers,sort&compress]{natbib}

\PassOptionsToPackage{numbers}{natbib}

\usepackage{color}
\bibpunct[, ]{[}{]}{,}{n}{,}{,}% Citation support using natbib.sty
\renewcommand\bibfont{\fontsize{10}{12}\selectfont}% Bibliography support using natbib.sty
\makeatletter% @ becomes a letter
\def\NAT@def@citea{\def\@citea{\NAT@separator}}% Suppress spaces between citations using natbib.sty
\makeatother% @ becomes a symbol again

\theoremstyle{plain}% Theorem-like structures provided by amsthm.sty

\theoremstyle{definition}

\theoremstyle{remark}

\begin{document}
%\articletype{ARTICLE TEMPLATE}

\title{Orchestration of Heterogeneous Experimental Machines via ROS2 for Automated Bulk Intermetallic Synthesis}

\author{\name{Wei-Sheng Wang\textsuperscript{a,b,$\dagger$}, Kensei Terashima\textsuperscript{a,$\dagger$}\thanks{$\dagger$ These authors contributed equally}, and Yoshihiko Takano\textsuperscript{a,b}}
\affil{\textsuperscript{a}National Institute for Materials Science, Tsukuba, Japan; \textsuperscript{b}University of Tsukuba, Tsukuba, Japan}
}

% orcid Wei-Sheng Wang 0009-0001-3572-5736

\maketitle

\begin{abstract}
    With advances in informatics applied to materials science, predicting the physical properties of numerous materials has become increasingly feasible, creating a growing demand for their experimental validation. It has been expected that the integration of robotic systems into experimental materials science excels at efficiently performing repetitive and time-consuming tasks without the need for human intervention, thus significantly increasing throughput and reducing the risk of human error, while there have been a limited number of reports tackled the synthesis process of solid bulk material so far possibly because of the complex as well as a wide variety of processes to deal with. In this paper, we report an automated arc melting system controlled by a robot operating system 2 (ROS2). Taking advantage of ROS2, we have constructed a machine that can handle multiple experimental apparatuses simultaneously with flexibility for future expansion of functions.  The constructed machine is capable of not only performing repeated operation of a specific process but also dealing with multiple elements for synthesis of intermetallic compounds. The system is expected to accelerate experimental validation of data-driven materials exploration.
\end{abstract}


\section{Introduction}
Automation and robotics have been highly effective in industry, where they enable large-scale production of standardized goods. In contrast, their application in scientific research has been more challenging, as experiments often require diverse, adaptable, and specialized approaches rather than uniform outputs\cite{ramprasad2017machine, rajan2015materials}. In recent years, advances in robotics and machine learning have made it possible to meet these requirements in various fields\cite{li2021recent}, and automated systems has started assisting researchers efficiently explore the vast combinatorial space needed for experiments\cite{butler2018machine, macleod2020self, masubuchi2018autonomous, MatsudaLiO2, negahdary2025automated}. For instance, such approaches have enabled researchers at the University of Liverpool to utilize robotics to establish an autonomous laboratory that works continuously for a long period of time\cite{burger2020mobile}, and in the thin-film domain, Bayesian optimization has been combined with robotics to accelerate the study of improving the conductivity of materials\cite{shimizu2020autonomous}. Following the aforementioned liquid and thin-film materials, solid-state material synthesis has started very recently to benefit from these technologies\cite{szymanski2023autonomous, ASTRAL}.

Despite the large demand for automatization, it remains difficult to achieve such an automated laboratory system especially regarding bulk materials. This could be partly because of the fact that most of the experimental machines are not designed to communicate with each other, and they are not ready for plug-and-play as there is a lack of unified system or guidelines to integrate them.  To address this gap, a system is required that enables us to build a flexible and stable automated experimental platform\cite{stein2019progress}.

This article introduces our approach of building such an automation platform. For hardware, we have chosen arc-melting as an automated synthesis process, which is one of the most common methods to obtain bulk intermetallic compounds. For software, we have adopted Robot Operating System 2(ROS2)
to ensure double flexibility of the platform\cite{bonci2023robot, erHos2019ros2, ye2023ros2}, namely (i) the ability to adapt to various types of experimental machines, and (ii) capability to integrate additional devices or functionalities in the future\cite{macenski2022robot}.  The modular and distributed architecture of ROS2 helps us to construct a platform that can adapt to the diverse requirements of experimental setups, even if the experimental equipments originate from different manufacturers. Moreover, ROS2 facilitates seamless communication between various robotic components and ensures scalability as the system evolves. The platform integrates multiple sensors, actuators, and robotic manipulators, all coordinated under a unified control framework. By asynchronous movements of each component controlled by ROS2, the constructed machine is capable of controlling sequential but parallelized motions, which is useful for achieving automation of scientific experimental operations such as materials synthesis with high flexibility for examining a vast variety of the combination of raw materials.

\begin{figure}[!h]
\centering
      \centering
      \includegraphics[width=\linewidth]{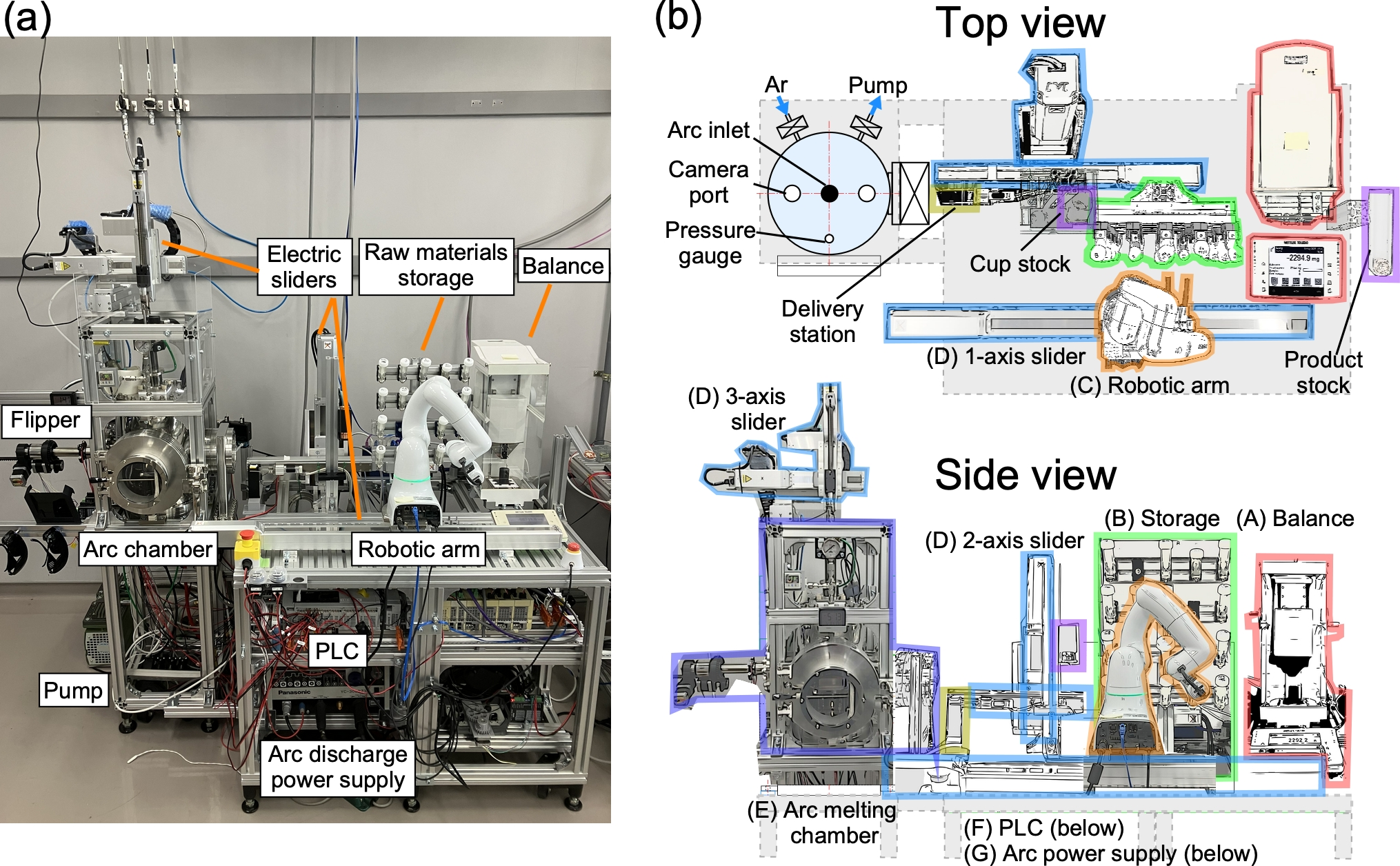}
      \caption{(a) A photograph of constructed automatic arc-melting system. (b) Schematic graph of the constructed system from top and side views with label for each apparatus.}
      \label{system hardware}
\end{figure}

\section{Experimental methods}
Stoichiometric ratio of vanadium (Nilaco, 99.7 $\%$) and germanium (Rare Metallic, 99.97 $\%$) are weighed to be 0.5 g and arc melted using the constructed machine with 60 A arc discharge power and 4 times of melting-flipping to ensure the homogeneity of samples.  Powder x-ray diffraction pattern of samples were measured using MiniFlex600, Rigaku.  The magnetic susceptibility measurements of samples under 10 Oe have been performed by MPMS XL-5, Quantum Design, in a zero-field cooling procedure.

\section{Constructed automatic arc melting system: Hardware}
The synthesis process of arc melting can be divided into three main factors, namely (i) weighing of raw materials, (ii) sample (or raw materials) transfer between apparatuses, (iii) arc melting operation including gas exchange in the arc chamber to Ar atmosphere, application of arc discharge to the sample, and several times of sample flipping and arc discharge loop to ensure homogeneity.  To automate these process, we have combined several machines as shown in Figure 1(a).  Figure 1(b) shows the schematic drawing of Fig. 1(a) from top view and side view, with label for each apparatus.  As in the figure, the system mainly consists of:
\def\theenumi{\Alph{enumi}}
\def\labelenumi{(\theenumi)}
\begin{enumerate}
    \item Electric balance with automated dispensing function (Mettler-Toledo, Quantos)\cite{bahr2018collaborative,bahr2020recent}
    \item Raw materials storage with dosing heads (Mettler-Toledo, QH012 series)
    \item 6-axis robotic arm (Denso-Wave, Cobotta)
    \item Electric sliders, 1-axis$\times$1, 2-axis$\times$1, 3-axis$\times$1 (IAI, RCS and PCP series)
    \item Vacuum chamber with arc discharge guide (Eurosystem), with user-attached water-cooled sample holder base, vacuum pump, electric gate valves, motored flipper, pressure gauge with vacuum switch (SMC), Water leak detector (Tatsuta, AD-AS-1WM), and a USB-camera for recording the synthesis process in the arc chamber (ANMo Electronics, Dino-Lite)
    \item Programmable logic controller (PLC) (Mitsubishi, FX5 series)
    \item Arc discharge power supply (Panasonic, YC-300BZ3)
\end{enumerate}

The standard operation process for this machine is as follows (see supplemental video).  The automated process begins with the 6-axis robotic arm and the 1-axis electronic slider transferring the copper sample carrier and dosing head containing raw material to the electronic balance. The raw material is dispensed on the sample carrier and its amount is weighed to the required quantity, after which the dosing head is returned to the storage. This weighing sequence is repeated until all components are dispensed on the copper sample carrier. Subsequently, the loaded sample carrier is transported via the robotic arm and slider to a delivery station adjacent to the arc chamber. The carrier is then moved into the arc chamber using a 2-axis electronic slider, after which the gate valve is closed. The chamber is evacuated and repeatedly purged with argon gas 4 times to ensure an inert atmosphere prior to arc-melting operations.  Once the purging is complete, the arc discharge is initiated and its trajectory is controlled by a 3-axis electric slider grabbing the arc-discharge guide.  After each arc-discharge, the sample is flipped by a motorized flipper.  Upon completion of the pre-set cycles of melting and flipping, the chamber is vented with Ar gas, and the gate valve opens. Then the 2-axis slider transfers the copper sample carrier containing the synthesized material from the arc chamber to a delivery station.  Upon arrival, the robotic arm, assisted by the 1-axis slider, grabs the copper holder and places it onto the electronic balance to record the total weight of the product sample.  Finally, the robotic arm transports the sample carrier from the balance to the stock position and returns to its initial position.  In the following subsections, more details are given for the main 3 factors of automated arc melting hardware: weighing of raw materials, sample transfer, and arc melting.

\subsection{Weighing of raw materials}
Manual weighing of raw materials is one of the most time-consuming steps in the synthesis process, as it requires careful attention to prevent contamination.  To automate this process while minimizing the risk of contamination, we employed a balance equipped with automatic weighing function.  This machine integrates the electronic balance with an automatic feeder head that dispenses material until the target weight is reached. By simply replacing the feeder head containing a specific raw material of individual atomic element, multiple substances can be weighed sequentially without cross-contamination. The equipment has been previously used in other dispensing targets \cite{lee2007optimisation,ota2021evaluation}, such as liquids and fine oxide powders. However, it has been found that weighing metallic raw materials for arc melting requires additional considerations.

First, the use of fine powders (typical particle size less than 100 ${\mu}m$) may be avoided, as they tend to disperse or splash when exposed to the electric arc.  Second, metallic fine powder materials sometimes cause stucking of the dispensing hole, defining the lower size limit.  Third, the raw materials need to be smaller than the dispensing hole (approximately 2 mm in diameter); even though the dosing heads with large dispensing hole is available, the fine control at the milligram scale becomes challenging with large pieces, thereby establishing the upper size limit.  Due to these constraints, we use coarse powders or small chunks (typical size of 1 mm in diameter) as raw materials and corresponding dosing head (QH012).  A detailed list of materials and their size in use is provided in the supplemental information.  The raw material storage can host a maximum of 20 dosing heads by 5 rows and 4 columns.

%{\color{red} 16 atoms are verified to work successfully with the constructed arc-melting system, that is, $B, Al, Si, Ti, V, Cr, Mn, Fe, Co, Ni, Cu, Ge, Zr, Nb, Mo, Ta$.}.

\subsection{Transfer of raw materials and product specimen}
To cover the distance required for transferring raw materials in the arc melting process, we have coupled a 6-axis robot arm with electric sliders. One of the possible configurations was to arrange all machines centered at the robot arm so that single robot can reach all, such as operated in Maholo\cite{yachie2017robotic}. On the other hand, considering our future system expansion, here the apparatuses, namely the raw material storage, the weighing balance, and the vacuum chamber for arc melting, are aligned in a straight manner. Although a multiple-axis robot arm can mimic human hand movements and is adaptable to existing systems, this linear arrangement exceeds the reach of a single arm, necessitating the use of additional sliders.

For a 6-axis robotic arm, we chose a collaborative one that does not require protective barriers, to easily ensure a safe working environment where the system needs human assistance in the construction or trials $\And$ errors period, as well as continuous maintenance and refilling of consumables. At the tip of the robot arm, we have designed and attached a 3D-printed gripper that can grab both the dosing heads and the sample carrier.

For electric sliders, we chose the ones with simple linear motion that allow us to easily plan their arrangement. Compared to installing multiple robotic arms, this approach significantly reduces costs while improving transfer efficiency.  A 1-axis slider is used to transport the 6-axis robot arm, while a 2-axis slider is used to transfer the copper sample holder from a delivery staion outside the chamber to the arc-discharge position inside arc chamber.

\subsection{Arc melting}
The arc melting chamber was newly designed based on a commercially available tee with ICF204 ports.  The chamber is required to be vacuum-proof since the arc melting has to be performed in Ar atmosphere with a weak vacuum of approximately 0.5 atm while the chamber pressure needs to be 1 atm for sample transfer.  Thus, the chamber was designed as small as possible to enable quick purging while maintaining the required ports for gas exchange and water cooling of arc discharge stage.  As a result, the constructed chamber could complete the purging process of 4 times of alternate pumping and Ar filling in less than 3.5 minutes, without a clear signature of oxidized byproduct during arc melting.  As in Figs. 1 and 2, the top of the chamber is connected via electrical insulation to an arc discharge guide with a celium-doped tungsten tip, which is constructed as a double pipe with internal water cooling.  The guide is connected to an arc discharge power supply, and also connected via electrical insulation to a 3-axis electric slider so that the arc discharge position can be controlled by manipulating the slider.

\begin{figure}[!h]
  \centering
      \centering
      \includegraphics[width=\linewidth]{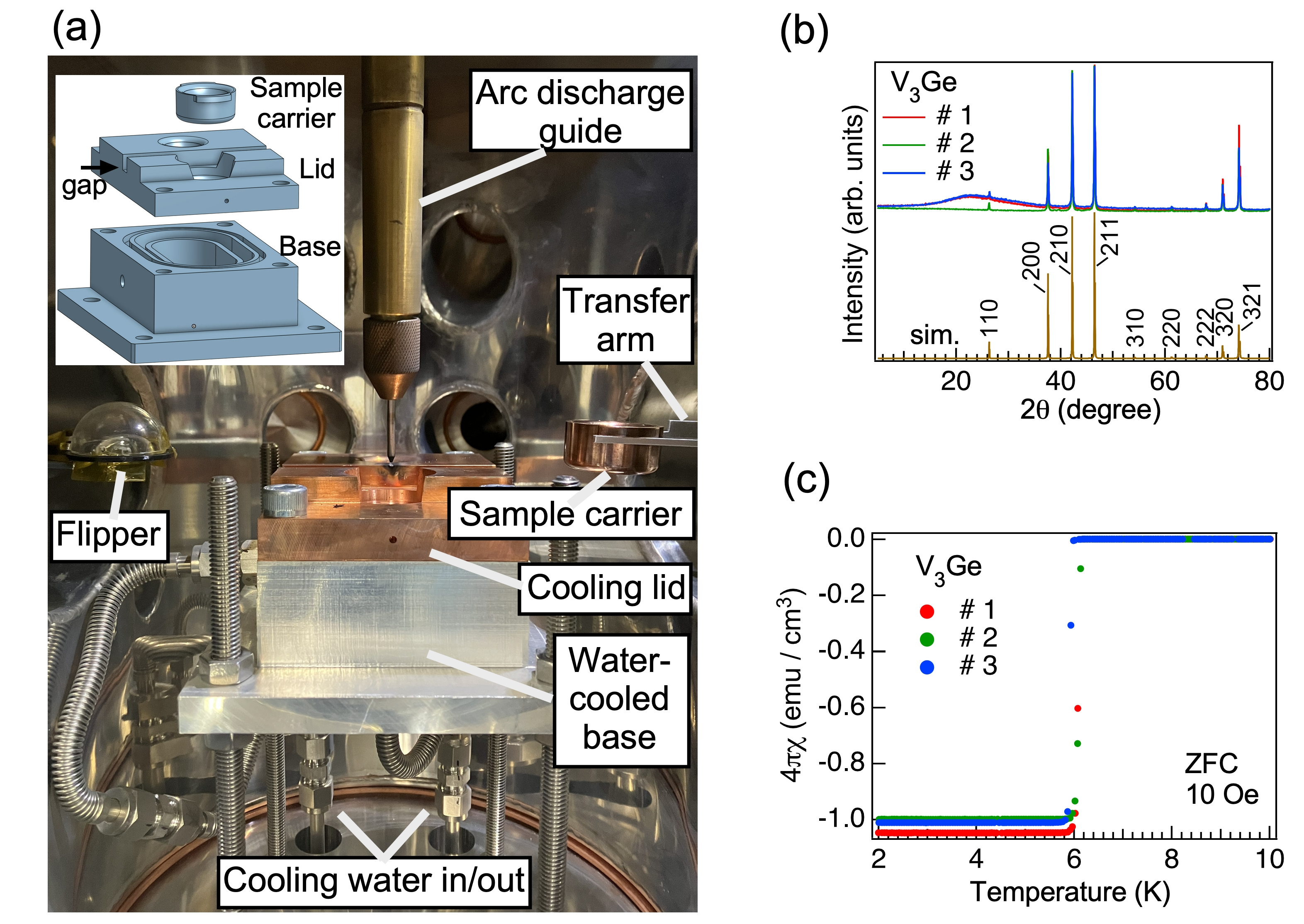}
      \caption{(a) Photograph of inside arc-melting chamber. Inset shows the detail of arc discharge stage drawn by a 3D CAD software OnShape.  (b) XRD patterns of synthesized V$_{3}$Ge samples compared with simulated patterns using VESTA\cite{VESTA}. (c) Magnetic susceptibility of synthesized V$_{3}$Ge samples taken by zero field cooling process at 10 Oe.}
      \label{arc base}
\end{figure}

Inset of Fig. 2(a) shows the detail of arc discharge stage.  The copper sample carrier serves as the primary vessel for holding the prepared raw materials. The copper cooling lid has two grooves, one for zirconium oxygen trap, and the other for the sample carrier.  The cooling water enters the stainless steel water-cooled base from the bottom and hits the cooling lid, then it is evacuated through the side of the water-cooled base.  A small vent hole is drilled in the lid to avoid trapping air between the sample carrier and the cooling lid. As indicated by an arrow in the inset of Fig. 2(a), there is a gap between the sample carrier and the cooling lid, which is to prevent the accident of co-melting and uniting them. The gap also provides the required space for the transfer arm to access and handle the sample carrier.  The sample carrier has a half-sphere shape dip to store the raw material or the product sample. The arc discharge starts from zirconium oxygen trap, and it approaches to the sample carrier with preset trajectories. When the arc discharge is over, 2-axis motorized flipper moves to the top of the sample carrier and rotates approximately 1 mm above the sample carrier to flip the sample for ensuring homogeneity of the sample.  The whole process of arc discharge and flipping is recorded by a USB camera.  The constructed arc melting process has several parameters in control: input ratio and amount of raw materials, the weighing order, the trajectory of arc discharge, the number of melting-flipping times, all of which are recorded. This provides a well-defined and digitalized as well as reproducible synthesis motion that helps the exploration of new materials.

Figures 2(b) and 2(c) show the powder x-ray diffraction (XRD) patterns and the result of magnetic susceptibility measurements of a well-known superconductor $V_{3}Ge$ \cite{Matthiasreview, Nakahira_2021, ahmed2022superconducting} with $T_c \sim 6 K$, synthesized by the constructed arc melting system.  The results of 3 samples overlap well each other, indicating that the well-controlled synthesis condition can be a powerful tool for exploration of functional intermetallic compounds.  All observed XRD peaks could be identified with those of the target $V_{3}Ge$ without a clear signature of the impurity phase. Assuming a demagnetizing field effect of a sphere, the estimated magnetic susceptibility of the samples was close to $-1/4\pi$ that corresponds to 100$\%$ shielding fraction.

\section{Constructed automatic arc melting system: Software and Control}
In this section, we describe the implementation of a flexible control system for hardware components from multiple manufacturers with different communication protocols and input/output (I/O) interfaces. To address the heterogeneity of the hardware and ensure seamless integration, we developed a control architecture based on ROS2 as it offers a modular, distributed framework that supports real-time communication, device synchronization, and scalable system integration\cite{bonci2023robot, erHos2019ros2, ye2023ros2}.

\subsection{Physical communication network}
The constructed system consists both of the machines possessing controller and communication port, and those without them.  The former can be directory connected to the main PC, while the latter needs an additional item to communicate.  As shown in the left panel of Fig. 3, in our case the former corresponds to electrical balance, robotic arm, electrical slider, USB camera, and secondary PC for storing synthesis records with electrical lab notebook service, which are connected to the main ROS2 operation PC by either LAN cable or USB. Whereas, the latter corresponds to the arc discharge power supply, electric gate valves, vacuum pump, sample flipper motor, and sensors such as limit switch, pressure gauge, water flow monitor and water leak detector. To bridge the communication between these components and the computer, we implemented a PLC.  The PLC performs essential automation tasks for such components through user-defined ladder programs, including automated vacuum pumping and venting sequences, as well as manipulating 2-axis motors for sample flipper. In addition, the PLC continuously monitors parameters derived from sensors such as vacuum levels and cooling water flow, ensuring safety during each step of the arc melting process.

\begin{figure}[!h]
\centering
  \includegraphics[width=\linewidth]{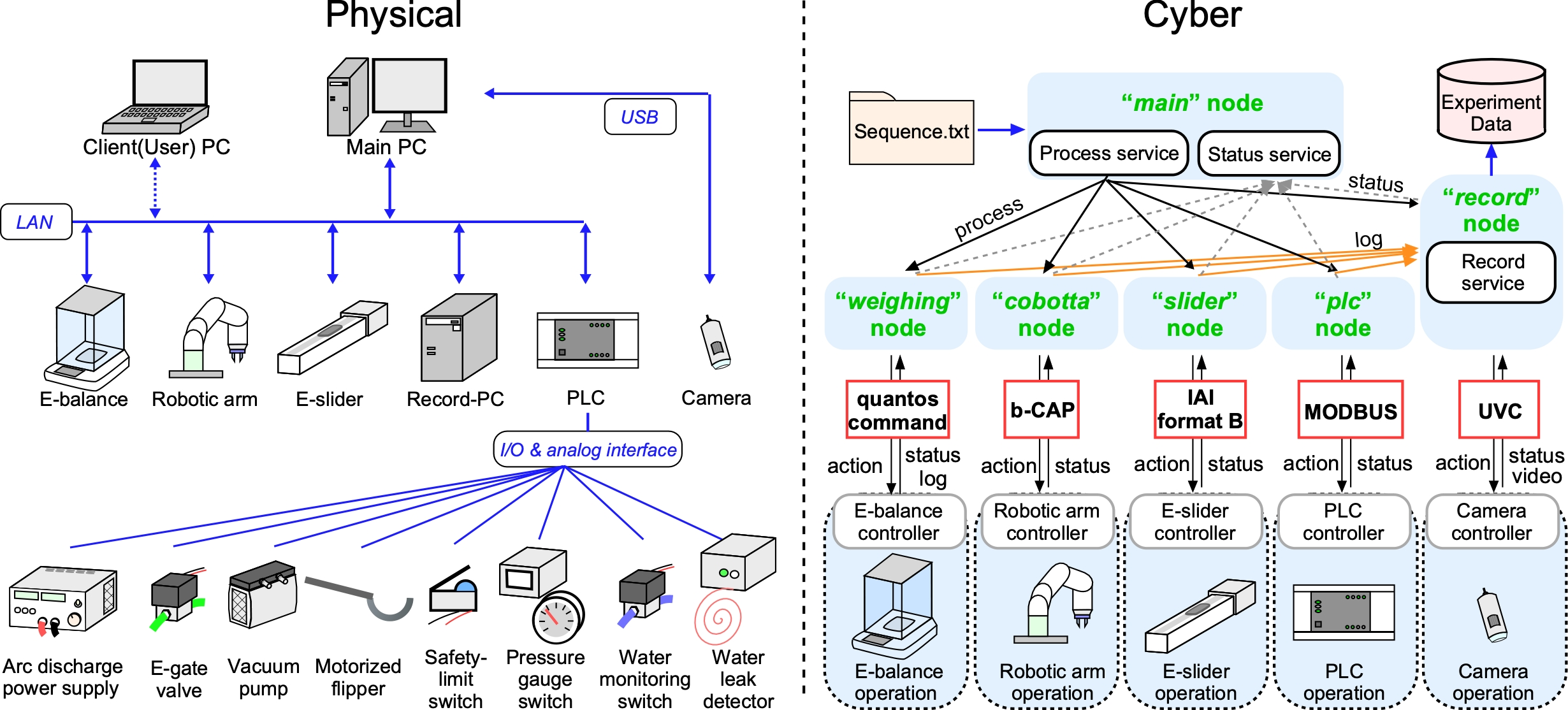}
  \caption{Physical (left) and cyber (right) communication network between apparatuses.}
  \label{system_communication}
\end{figure}

\subsection{ROS2-based operation software}
The integration of a PLC enabled the construction of a system in which all hardware possess individual controllers that can receive commands from the main program to perform various tasks. To obtain a unified system for devices from different manufacturers, we utilized ROS2 with a weak-coupling system framework\cite{bonci2023robot, erHos2019ros2, ye2023ros2}, where each component works independently before integrating it into the overall system so that failures or modifications in one part do not affect the overall system. This approach facilitates the modular unit with easy debugging and allows the seamless integration of new equipment into the system in the future, as we show the general workflow at Section\ref{workflow}. Additionally, the ROS2-based system architecture supports asynchronous execution, enabling multiple devices to operate concurrently and to avoid time bottleneck in a series of sequences, as we discuss in detail at Section\ref{comparison}.

The architecture of the developed ROS2-based software system is schematically illustrated in the right panel of Fig. 3. The system comprises six key functional ROS2 nodes: $main, weighing, cobotta, slider, plc$, and $record$. The $main$ node orchestrates the overall process, while the $weighing, cobotta, slider$, $plc$, and $record$ nodes each interface with their respective hardware controllers.

The $main$ node sequentially publishes process messages defined in $Sequence.txt$, starting from the first line. Each process message may contain one or more predefined $nodename\_parameter1\_parameter2\_...$ strings (hereafter abbreviated as $nodename\_action$), such as $plc\_pump$, which instructs the $plc$ node to initiate the pumping procedure. Upon detecting its respective $nodename\_action$ in the process message, a node issues the corresponding command to its hardware controller in accordance with the controller’s communication protocol.  This approach enables the system to provide uniform control of heterogeneous machines originating from different manufacturers with varied communication protocols. Simultaneously, the $Status$ service in the $main$ node is listening to the status messages uploaded by all other nodes. The definition of status is given in each node depending on the response from each controller, and they are defined as "action", "standby", and "error". "Status" service in the main node is monitoring the status of all nodes, and it sends a signal to "Sequence" service of moving to next step in the sequence when all nodes report a standby status. Multiple $nodename\_action$ strings within a single line of $Sequence.txt$, separated by spaces, enable simultaneous operation of different nodes, thereby supporting asynchronous execution. For example, the sequence
$"plc\_pump$ $slider\_weightPos$ $weighing\_open"$
initiates pumping of the arc chamber via the $plc$ node, moves the 6-axis robotic arm in front of the electronic balance using the $slider$ node, and opens the balance’s front door through the $weighing$ node concurrently.  For execution of a new synthesis experiment, users do not need to modify any ROS2 node program, but only need to prepare the corresponding $Sequence.txt$ before running the $main$ node from their client PC. To perform a standard synthesis process specified in Section 3, the users can also use a GUI-based auxiliary Streamlit \cite{streamlit} program also provided in the github that generates $Sequence.txt$ by specifying the input parameters, namely the composition of the target material, the desired weight, the weighing order, the number of melting-flipping times and the trajectory of arc discharge.

The $record$ node subscribes to the $main$ node and logs the sequence message with timestamp. In addition, the $record$ node has a recording service that listens to other nodes so that specified messages can be saved to the log, for example, the raw material weights reported by the $weighing$ node. Furthermore, the $record$ node saves the video footage of the arc-synthesis experiments on request. All system logs are stored locally and concurrently uploaded to a remote instance of eLabFTW—an open-source electronic laboratory notebook\cite{hewera2021elabftw}. This dual logging strategy ensures redundancy and facilitates the traceability of the experimental workflow.

\begin{figure}[!h]
\centering
  \includegraphics[width=\linewidth]{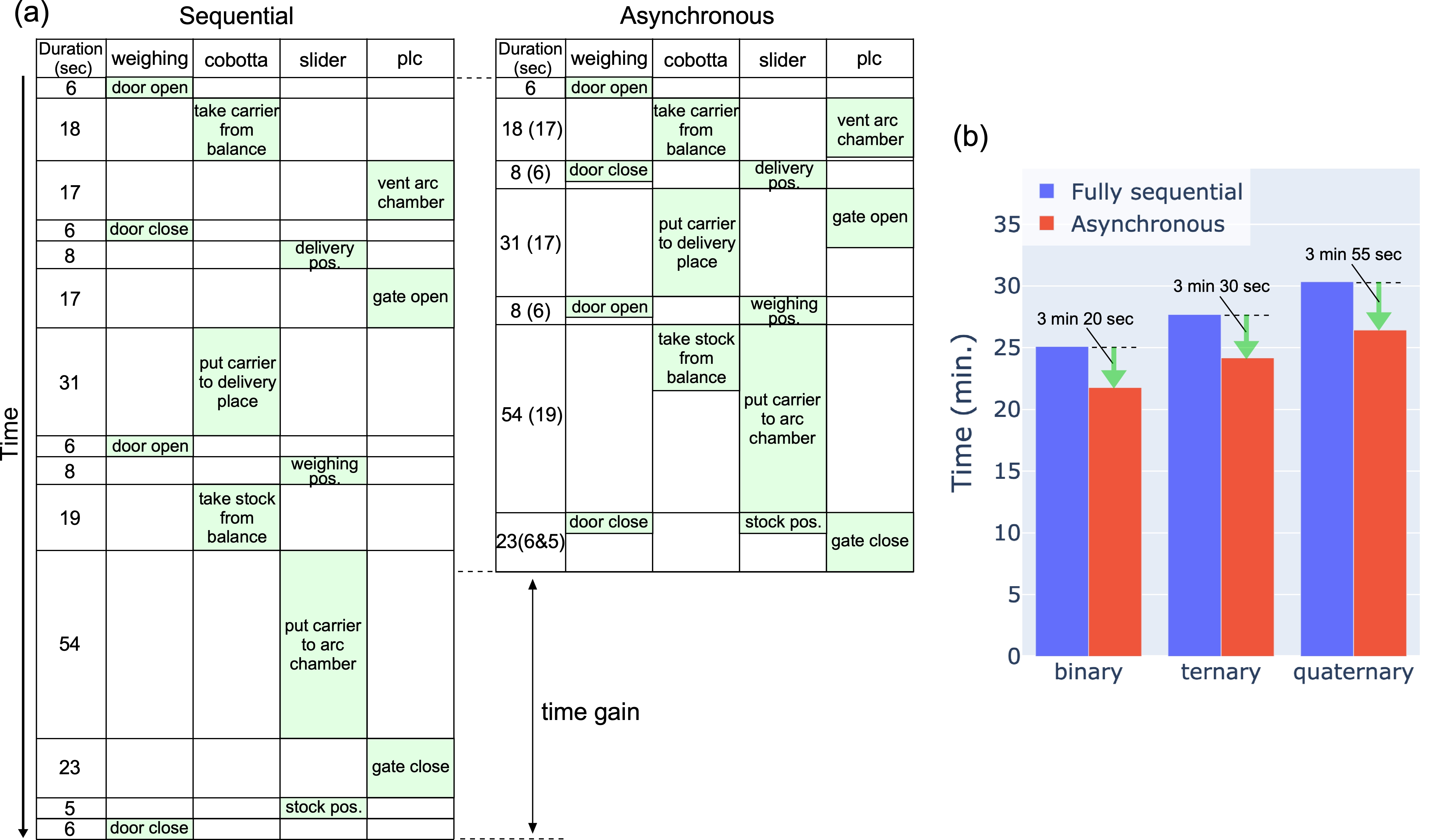}
  \caption{(a) A schematic comparison of time duration between fully sequential and asynchronous process control, using a part of actual process described in Section 3 as an example. For a detailed duration of each process, see supplemental video. Numbers in parenthesis at asynchronous table show the time duration for the sequence done simultaneously, corresponding to the earned time. (b) Comparison of time required to complete binary, ternary and quaternary inter-metallic compound synthesis processes, between fully sequential and asynchronous operations.}
  \label{system_communication}
\end{figure}

\subsection{Comparison between fully sequential control and ROS2-based asynchronous operation}
\label{comparison}

The constructed weakly coupled framework of the ROS2 system allows each hardware component to be operated independently as a node\cite{bonci2023robot, erHos2019ros2, ye2023ros2}, enabling asynchronous yet sequential process orchestration. This approach is particularly suitable for automating synthesis workflows, where certain operations must be executed only after preceding tasks are completed or require waiting for ongoing procedures to finish, many of which are time-consuming. Parallelization of such processes can significantly improve overall efficiency, as schematically illustrated in Fig. 4(a).

In our case, the initialization process, which brings each machine into a safe posture and standby state, can be performed independently and simultaneously. Furthermore, the system is programmed such that, while the arc melting chamber is in motion, the six-axis robotic arm and slider simultaneously return the last-weighed raw material stock from the balance to the storage position. Figure 4(b) compares the execution times of the standard arc melting sequence described in Section 3. The plotted durations represent the average of three trials for each condition: binary (0.4 g + 0.1 g Al), ternary (0.3 g + 0.15 g + 0.05 g Al), and quaternary (0.3 g + 0.1 g + 0.05 g + 0.05 g Al), each taken from different stock positions and arc-melted 4 times (and flipped 3 times) in the arc chamber. The detailed experimental logs for representative trials and time durations for each trial are in the supplemental information.  The results clearly demonstrate that asynchronous operation with ROS2 exhibits higher efficiency, particularly as the number of processes increases and the operations become more complex. These findings indicate that asynchronous, parallelized operation can substantially accelerate sophisticated materials science experiments.

% \begin{figure}[!h]
% \centering
%     \begin{minipage}{0.2\textwidth}
%       \centering
%       \includegraphics[width=\linewidth]{photo/Nb3Al_P4_sample.png}
%       \caption{Sample.}
%       \label{system}
%     \end{minipage}
%     \begin{minipage}{0.01\textwidth}
%       \includegraphics[width=\linewidth]{photo/white.png}
%       \centering
%     \end{minipage}
%     \begin{minipage}{0.34\textwidth}
%       \centering
%       \includegraphics[width=\linewidth]{photo/Nb3Al_P4_XRD.jpg}
%       \caption{XRD.}
%       \label{arc}
%     \end{minipage}
%     \begin{minipage}{0.01\textwidth}
%       \includegraphics[width=\linewidth]{photo/white.png}
%       \centering
%     \end{minipage}
%     \begin{minipage}{0.4\textwidth}
%       \centering
%       \includegraphics[width=\linewidth]{photo/Nb3Al_P4_SQUID.png}
%       \caption{SQUID.}
%       \label{squid}
%     \end{minipage}
% \end{figure}

\section{General workflow for automating the experimental machines}
\label{workflow}

Our approach combines ROS2 with PLC for automating existing scientific experimental apparatus, which is highly flexible and would be applicable to a variety of machines. In this section, we show a general workflow regarding how to apply the proposed concept to the reseachers' own system, with an example to integrate new function into the system.  Our entire ROS2 codes will be available at github upon acceptance of the manuscript, where it contains a $template$.  The connectable candidate apparatuses might be divided into 2 kinds: (i) machines containing an independent controller with LAN communication, and (ii) sensors and machines without LAN communication.  The former can be a new ROS2 node, while the latter needs additional device such as PLC to communicate as we will describe in detail below.

\begin{figure}[!h]
\centering
  \includegraphics[width=\linewidth]{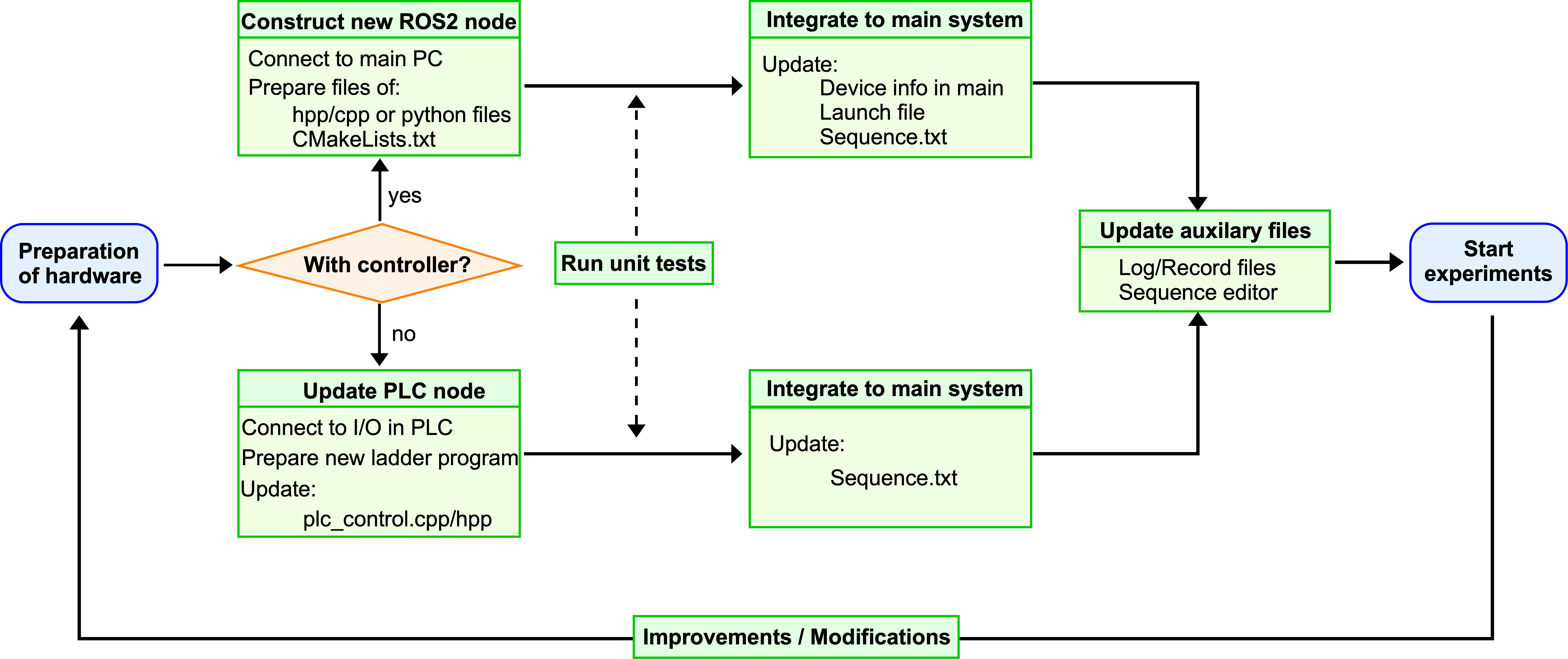}
  \caption{A schematic workflow to automate experimental apparatus.}
  \label{system_communication}
\end{figure}

\subsection{Machines with built-in controllers and LAN connectivity}
If the apparatus already contains an independent controller but does not support LAN, a network adapter can be used to establish the LAN connection for system integration. Once the apparatus is LAN enabled, it indicates that the device has a certain level of automation and command execution capability. At this stage, it is necessary to clarify the communication protocol and the command structure of the device.

\begin{itemize}
    \item If the manufacturer provides a library, one may directly use it to implement a ROS2 node in \texttt{C} or \texttt{Python}.
    \item If the communication protocol needs to be defined manually, the developer may choose whichever programming language is most convenient.
    \item For consistency in workflow design, we recommend adopting a command format such as: 
    $NodeName\_Parameter1\_Parameter2\_...$, which allows for structured command generation and easier maintenance.
\end{itemize}
After implementation, each communication command should be verified through unit testing to ensure proper execution.

\subsection{Machines without built-in automation}
If the apparatus does not provide any automation interface, the first step is to make it controllable through power control:

\begin{itemize}
    \item For example, a vacuum pump can be switched on/off by driving the motor switch mechanically, or more conveniently by using a relay to directly control its power supply.
    \item If the apparatus can only be controlled through power switching and does not provide LAN or an internal controller, an additional controller such as a microcontroller, Arduino, or Raspberry Pi can be introduced for integration as a new node.
\end{itemize}

In this context, the PLC can be a reliable controller as it is robust for long-term operation with LAN interfaces. After confirming that the apparatus can be operated through the PLC, the next step is to program the PLC logic (so called ladder) to control desired apparatus operations. The subsequent procedure is similar to case (i), where a ROS2 node is implemented to interface with the system.

\subsection{Unit testing and verification}
After the apparatus is successfully connected and controlled, unit testing should be performed to ensure stability and reliability before integrating into the full workflow in the following steps:

\begin{enumerate}
    \item Run the ROS2 node to test, and examine whether it reacts properly in a single execution by sending program arguments. For example, use $ros2\ run\ slider\ init$ for $slider$ node to send an action message that moves the slider to the position defined as $init$.
    \item Run the ROS2 node to test, and try if the same operation in (A) can be performed through the $main$.  For example, $Sequence.txt$ can be $slider\_init$ only.
    \item Once all individual commands have been confirmed to work as intended, the apparatus can then be integrated into the overall workflow for system-level testing.
\end{enumerate}

\section{Summary}
This study introduced an automated synthesis platform for bulk intermetallic compounds based on an arc-melting process. By integrating heterogeneous experimental apparatuses under a ROS2-based control system, we achieved a fully automated workflow that improves reproducibility and efficiency. The system’s flexibility allows seamless addition of new devices and experimental functions, while asynchronous operation reduces bottlenecks. Validation through successful synthesis of superconducting V$_3$Ge points to the potential of this platform to accelerate data-driven discovery of functional materials.

%Our automated synthesis platform has already demonstrated the capability to systematically collect data. To fully leverage this system, we aim to expand its scope to include the synthesis of complex alloys and advanced functional materials. This will generate a larger dataset, enabling machine learning algorithms to uncover previously unknown correlations and optimize the synthesis process more effectively.
%Developing a modular and scalable automated platform is crucial for long-term success. Collaboration among academia, industry, and technology developers will be essential to overcome current limitations and realize the vision of autonomous experimentation. Such efforts will bridge the gap between aspiration and realization, ultimately revolutionizing the process of scientific discovery.

%We envision a future where automated platforms not only support but actively lead scientific exploration. These advancements will accelerate material discoveries, significantly benefiting society and driving innovation across various disciplines.

\section*{Acknowledgment}
This work was supported by supported by JSPS KAKENHI (Grant Nos. 23K04572, 23KK0088, 24K01333).  W.S.W. acknowledges the support by JST SPRING, Grant Number JPMJSP2124.
 
\bibliographystyle{unsrtnat}
\bibliography{references}  %%% Uncomment this line and comment out the ``thebibliography'' section below to use the external .bib file (using bibtex) .

%%% Uncomment this section and comment out the \bibliography{references} line above to use inline references.
% \begin{thebibliography}{1}

% 	\bibitem{kour2014real}
% 	George Kour and Raid Saabne.
% 	\newblock Real-time segmentation of on-line handwritten arabic script.
% 	\newblock In {\em Frontiers in Handwriting Recognition (ICFHR), 2014 14th
% 			International Conference on}, pages 417--422. IEEE, 2014.

% 	\bibitem{kour2014fast}
% 	George Kour and Raid Saabne.
% 	\newblock Fast classification of handwritten on-line arabic characters.
% 	\newblock In {\em Soft Computing and Pattern Recognition (SoCPaR), 2014 6th
% 			International Conference of}, pages 312--318. IEEE, 2014.

% 	\bibitem{hadash2018estimate}
% 	Guy Hadash, Einat Kermany, Boaz Carmeli, Ofer Lavi, George Kour, and Alon
% 	Jacovi.
% 	\newblock Estimate and replace: A novel approach to integrating deep neural
% 	networks with existing applications.
% 	\newblock {\em arXiv preprint arXiv:1804.09028}, 2018.

% \end{thebibliography}

\end{document}